%% file: SingleOPOSquareGrid_main.tex
\documentclass[prx,superscriptaddress,reprint,showpacs,amsmath,amssymb,nofootinbib%
,eqsecnum%
]{revtex4-1}

\include{preamble}
\begin{document}

\title{One-way quantum computing with arbitrarily large time-frequency continuous-variable cluster states from a single optical parametric oscillator}

\author{Rafael N. Alexander}
\affiliation{School of Physics, The University of Sydney, Sydney, NSW 2006, Australia}
\affiliation{School of Science, RMIT University, Melbourne, VIC 3001, Australia}
\author{Pei Wang}
\author{Niranjan Sridhar}
\author{Moran Chen}
\author{Olivier Pfister}
\email{opfister@virginia.edu}
\affiliation{Department of Physics, University of Virginia, Charlottesville, Virginia 22903, USA}
\author{Nicolas C. Menicucci}
\email{ncmenicucci@gmail.com}
\affiliation{School of Physics, The University of Sydney, Sydney, NSW 2006, Australia}
\affiliation{School of Science, RMIT University, Melbourne, VIC 3001, Australia}

\date{\today}

\begin{abstract}%
One-way quantum computing is experimentally appealing because it requires only local measurements on an entangled resource called a cluster state. Record-size, but non-universal, continuous-variable cluster states were recently demonstrated separately in the time %
and frequency domains%
. We propose to combine these approaches into a scalable architecture in which a single optical parametric oscillator and simple interferometer entangle up to ($3\times 10^3$~frequencies) $\times$ (unlimited number of temporal modes) into a 
new and computationally universal continuous-variable cluster state. We introduce a generalized measurement protocol to enable improved computational performance on the new entanglement resource.
\end{abstract}

\maketitle

\section{Introduction}
One-way quantum computing~\cite{Raussendorf2001} is a form of measurement-based quantum computing (MBQC)~\cite{Gottesman1999,Jozsa2005} and an appealing alternative to the circuit model~\cite{Nielsen2000},  which is being more widely pursued~\cite{Ladd2010}.  In one-way quantum computing, the primitives of the universal gate set are pre-encoded in a ``quantum substrate'' that is a generic, yet precise, entangled cluster state described by a graph specifying the entanglement structure of the qubits~\cite{Briegel2001} or qumodes~\cite{Zhang2006}. Quantum computing  proceeds solely  from single-node measurements on the cluster graph and feedforward of the measurement results%
 ~\cite{Raussendorf2001,Menicucci2006}.

Quantum error correction and fault tolerance in one-way quantum computing have been theoretically proven feasible for qubit cluster states~\cite{Aliferis2006}, with thresholds comparable to those for concatenated codes ($10^{-3}$ to $10^{-6}$), and then later improved using topological methods to thresholds slightly above the percent level~\cite{Raussendorf2006}. Fault tolerance has recently been proven for continuous-variable~(CV) cluster states in terms of required levels of squeezing, the squeezing threshold being no more than 20.5~dB for a $10^{-6}$ error rate~\cite{Menicucci2014}. Since the techniques used in Ref.~\citenum{Menicucci2014} mirror those in Ref.~\citenum{Aliferis2006},  this threshold value is conservative and can most likely be improved. 

A fully fledged, scalable experimental demonstration of one-way quantum computing has yet to be achieved, as none of the proof-of-principle implementations using four photonic qubits~\cite{Walther2005,Barz2012} or four optical qumodes~\cite{Ukai2011} employed a scalable architecture.

Recently, one-dimensional cluster-state entanglement was demonstrated, at record sizes, over the continuous variables represented by the quantum amplitudes of the electromagnetic field, a.k.a.\ qumodes.  This was achieved in the frequency domain~\cite{Chen2014}, with 60 simultaneously addressable entangled qumodes, and in the time domain~\cite{Menicucci2011a,Yokoyama2013}, with $10^{4}$ sequentially addressable entangled qumodes.  Solely  technical issues reduced these numbers from their potential higher values of $3\times10^{3}$ qumodes in the frequency architecture~\cite{Wang2014a} and unlimited qumodes in the temporal architecture~\cite{Yokoyama2013}. 
Besides this scalability breakthrough, optical implementations of quantum information offer other advantages such as room temperature operation, naturally low decoherence, and significant  potential for device integration~\cite{Politi2009, Bauters2011}.

In this paper, we show that one can create computationally universal CV cluster  states 
by  entangling, both in time and in frequency, the quantum frequency comb of EPR pairs emitted from a single  optical parametric oscillator~(OPO).  Based on previous results~\cite{Yokoyama2013,Chen2014}, the lattice for this state could potentially be up to $3\times 10^3$~nodes in one dimension (frequency) and unlimited in the other (time bins). We then show that this state enables universal quantum computing. 

\begin{figure*}[t!]
\centering
\includegraphics[width=1.8\columnwidth]{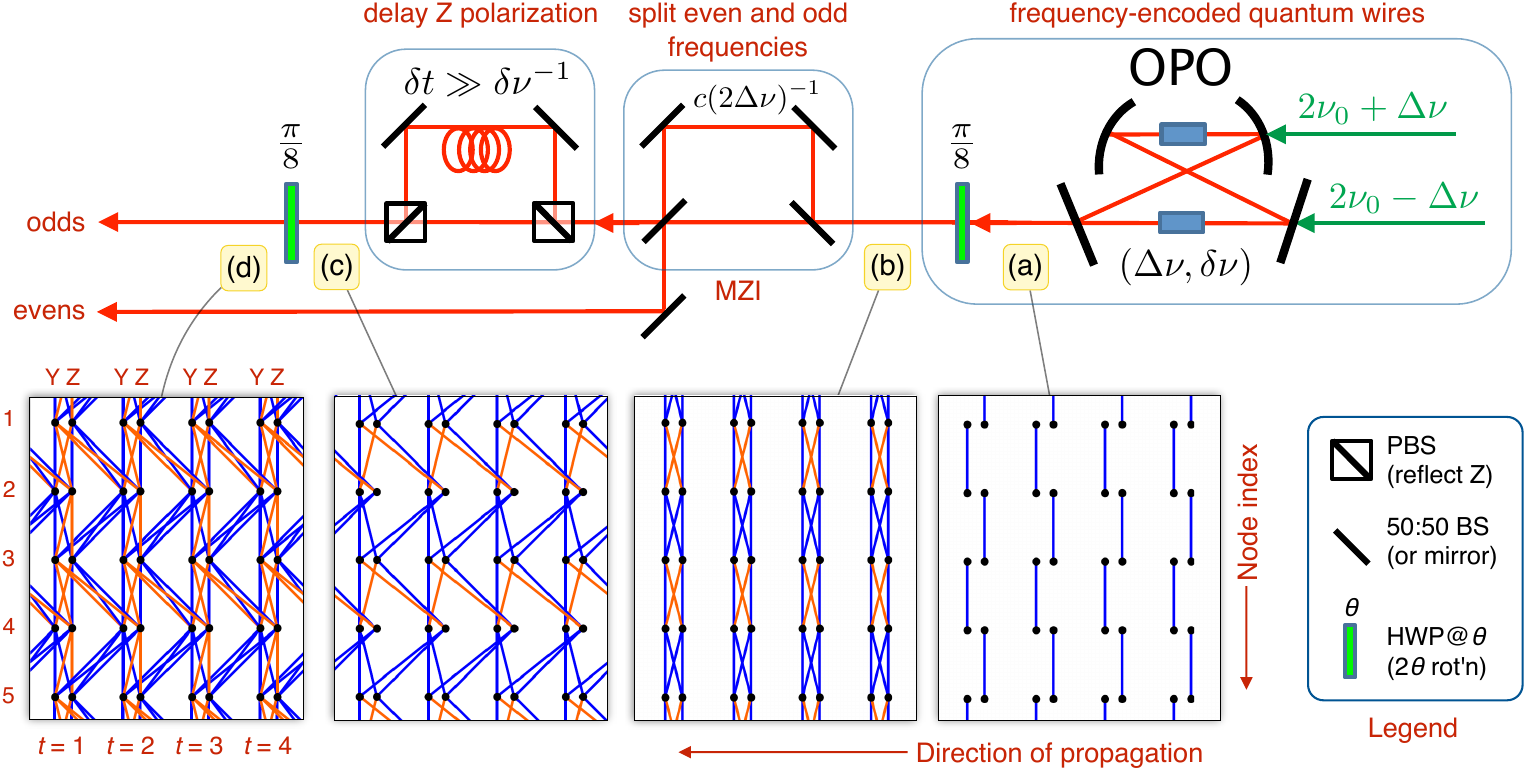}
\caption{%
Experimental setup to generate a bilayer square-lattice (BSL) CV cluster state (see text for details). Abbreviations: HWP@$\theta$ = halfwave plate at angle~$\theta$ to the horizontal principal axis of the crystal (rotates polarization by~$2\theta$); (P)BS = (polarizing) beamsplitter;  MZI~= Mach-Zehnder interferometer.  Local oscillator fields, at the frequencies of the qumodes to be measured, will be injected at the unused input port of the  MZI  and will also be used for locking all optical phases in the experiment. Note that light propagates from right to left in the figure. The labeled panels show a precise graphical representation of the Gaussian state present in the beam at each step of the experiment, using the simplified graphical calculus for Gaussian pure states (for notation and definitions, see %
Appendix~\ref{sec:introGC}).  Blue and orange correspond to edge weights of $\pm \mathcal C \tanh 2r$~\cite{Menicucci2011a}, respectively, with $\mathcal C$ given below for each panel. All qumodes (black dots) are labeled as shown in the left panel: by node index [Eq.~\eqref{main:eq:nodeindex}] (vertical) and by time bin and polarization (horizontal). \textbf{(a)}~The OPO generates a temporal sequence of frequency-encoded two-mode squeezed states ($\mathcal C=1$). \textbf{(b)}~Multiple (time-binned) CV dual-rail quantum wires encoded in frequency~\cite{Chen2014,Wang2014} ($\mathcal C = 2^{-1/2}$). \textbf{(c)}~Result of delaying all odd-numbered $Z$-polarized qumodes ($\mathcal C=2^{-1/2}$). \textbf{(d)}~Final BSL CV cluster state ($\mathcal C=2^{-3/2}$) after required phase  delays (see text).%
}
\label{main:fig:setupgraph}
\end{figure*}

This work combines the best of all previous proposals for \emph{scalable} CV cluster states: It employs Gaussian states with bipartite, self-inverse graphs---which are known to be highly scalable~\cite{Menicucci2008,Flammia2009}---and reduces the experimental requirements by simultaneously utilizing both frequency multiplexing~\cite{Chen2014,Wang2014} and temporal multiplexing~\cite{Menicucci2010,Menicucci2011a,Yokoyama2013}. In addition, these architectures are known to admit more compact computation~\cite{Yokoyama2013} with more favorable noise properties~\cite{Alexander2014} when compared to approaches based on CV cluster states generated by the canonical method~\cite{Menicucci2006,Gu2009}. Those so-called canonical CV cluster states~\cite{Menicucci2011}---which also admit a temporal~\cite{Menicucci2010} and a time-frequency implementation~\cite{Humphreys2014a}---are not so easily scalable in optics due to frequent use of the CV controlled-Z gate.%

Our proposal, in contrast, employs macronode-based cluster states~\cite{Flammia2009} entangled into a \emph{bilayer square lattice}~(BSL), which has 2 qumodes per macronode (hence `bilayer'), instead of 4 as in previous proposals~\cite{Menicucci2008,Flammia2009,Wang2014,Menicucci2011}. The BSL CV  cluster state admits a more versatile elementary gate set than do canonical CV cluster states~\cite{Menicucci2006,Gu2009}, generalizing an analogous result for single-qumode operations on the CV dual-rail quantum wire~\cite{Alexander2014}.%

The structure of this Article proceeds as follows: In Sec.~\ref{sec:stategen} we describe the BSL resource state and give an explicit experimental procedure for how to generate it. In Sec.~\ref{sec:verification} we discuss the experimental requirements in detail. In Sec.~\ref{sec:computing} we outline the basics of our measurement protocol. In Sec.~\ref{sec:univgateset} we describe how to implement universal quantum computation on the BSL. In Sec.~\ref{sec:noise} we discuss noise due to finite squeezing and we conclude in Sec.~\ref{sec:conc}. 
\section{State generation}\label{sec:stategen}

Construction of the BSL CV cluster state is illustrated in Fig.~\ref{main:fig:setupgraph} and described in more detail here. A type-II OPO is pumped at two frequencies~$2\nu_0 \pm \Delta\nu$,  one of each polarization ($Y$ and $Z$). Each pump produces a number of two-mode squeezed~(TMS) states~\cite{Ou1992a} over the frequency comb of the OPO eigenmodes, as shown in Fig.~\ref{main:fig:setupgraph}(a). These states are each a Gaussian approximation to an Einstein-Podolski-Rosen~(EPR) state~\cite{Einstein1935} between two frequencies that add to the corresponding pump frequency. Now, even if the pump beams are continuous wave, we still can, and will, logically assign pieces of the output beam to sequential time bins~\cite{Yokoyama2013}. 

The OPO modes have linewidth~$\delta\nu$ and are spaced by the free spectral range $\Delta\nu$. Each output frequency $\nu_n = \nu_0 + n\Delta \nu$ has a corresponding frequency index~$n$ and associated macronode index~\cite{Wang2014}
\begin{align}
\label{main:eq:nodeindex}
	m \coloneqq (-1)^n n\,,
\end{align}
which we will call the \emph{node index} for short and is used to label qumodes sequentially  (rather than by frequency)  in Fig.~\ref{main:fig:setupgraph}(a). Indeed, phasematching two frequencies~$\nu_n$ and~$\nu_{n^\prime}$ 
requires ${n + n^\prime} = \pm 1$,  and all TMS states are generated between adjacent node indices  (i.e., $m - m^\prime = \pm1$~\cite{Wang2014}) in Fig.~\ref{main:fig:setupgraph}(a).

A $\tfrac \pi 4$ polarization rotation  (by a halfwave plate at $\tfrac \pi 8$ rad from the horizontal principal axis of the OPO crystals), equivalent to a balanced beamsplitter for polarization qumodes, entangles these TMS states into a temporal sequence of frequency-encoded dual-rail quantum wires~\cite{Chen2014,Wang2014}, as shown in Fig.~\ref{main:fig:setupgraph}(b).
A Mach-Zehnder interferometer~(MZI) of path difference $c(2\Delta\nu)^{-1}$~\cite{Glockl2004,Huntington2005} separates frequencies of even and odd frequency index (and node index) into separate beams. For all odd qumodes, the $Z$ polarization is then time-delayed with respect to the $Y$ polarization by the interval $\delta t$ between two consecutive time bins. The result is shown in Fig.~\ref{main:fig:setupgraph}(c). A final $\tfrac \pi 4$ polarization rotation on the odd qumodes (another ``balanced beamsplitter'') yields the BSL graph of Fig.~\ref{main:fig:setupgraph}(d).

A final phase  delay by $\tfrac \pi 2$ (not shown) on either all odd or all even frequencies converts this into a finitely squeezed CV cluster state with the same ideal graph as in panel~(d). \emph{It is this state that we call the BSL CV cluster state.} The fact that the BSL is a \emph{bipartite, self-inverse graph} makes this possible and ensures the scalability of the scheme~\cite{Menicucci2008,Flammia2009,Menicucci2011,Menicucci2011a,Wang2014}.  
 (See the general discussion of bipartite, self-inverse graphs in Ref.~\citenum{Menicucci2011}.)

\section{Experimental Details}\label{sec:verification}

%

We now outline the  basic experimental requirements for generating the BSL CV cluster state, verifying its entanglement structure, and using it for quantum information processing.

\emph{Generating} the BSL CV cluster state requires a ``musical score'' condition---i.e., the measurement times must be compatible with resolving all qumode frequencies:\ $\delta t\gg\Delta\nu^{-1}$, an easily fulfilled condition. In addition, the measurement times must allow one to achieve maximum squeezing---that is, they must be at least as long as the OPO cavity storage time~\cite{Reynaud1987}. This translates into $\delta t\gg\delta\nu^{-1}\gg\Delta\nu^{-1}$, since  $\delta\nu$ is also half the squeezing bandwidth~\cite{Walls1994}. This condition can also be  easily  fulfilled in practice~\cite{Yokoyama2013} and ensures that the time bins contain maximally squeezed qumodes, to the extent permitted by the experiment's squeezing limit (mainly determined by the intracavity losses).%

Moreover, it is important to remember that as long as the undepleted pump approximation remains valid, the number of modes to be entangled has no bearing on the required pump power. 
To see that the undepleted-pump approximation holds for our scheme, note that a typical 100-mW pump power (i.e., ${2.5 \times 10^{17}}$~photons/s for green light) and a typical OPO cavity lifetime of 20~ns together yield ${5\times 10^{9}}$ pump photons available for downconversion. Squeezing of 20~dB corresponds to 24.5~OPO photons per output mode (since ${\avg {\op a^\dag \op a} = \sinh^2 r}$, with ${\#\text{dB} = 10 \log_{10} e^{2r} \approx 8.69\ r}$). With each pump photon downconverting into two daughter photons, even with one million output modes the total number of pump photons required is only $\frac 1 2 \times 24.5 \times 10^{6} = 1.2 \times 10^7$, which is just $0.25\%$ of the total number available. Therefore, pump depletion is indeed negligible.

%
%

%
To \emph{verify} that the BSL CV cluster state has been generated successfully,  we use a balanced homodyne measurement with a two-tone local oscillator (LO), as demonstrated in our two previous works~\cite{Pysher2011,Chen2014}. For entanglement characterization alone, the qumodes do not need to be separated in frequency.

%

When \emph{using} the BSL CV cluster state for quantum information processing, complete qumode separation is required. The qumode separation is straightforward in the time and polarization domains. Experimental techniques that have been honed on classical optical frequency combs~\cite{Diddams2007} can be used for the frequency domain qumode separation. Such techniques include virtually-imaged phase arrays (VIPAs), arrayed waveguide gratings (AWGs), as well as diffraction gratings and combinations thereof. After separation, the individual beams will be directed to homodyne detectors or photon counters as required by the particular algorithm~\cite{Gu2009}.
In the case of homodyne detection, the local oscillators will likely need to be derived from a stable classical comb, be it a femtosecond laser or a cavity-enhanced EOM, whose beam can be overlapped with the OPO's and subjected to the same frequency separation method. The use of integrated optics 
may assist in
implementing this scheme to full scale. %

%
\section{Basics of quantum computing on the bilayer square lattice}\label{sec:computing}

\begin{figure}
\includegraphics[width=1.0\columnwidth]{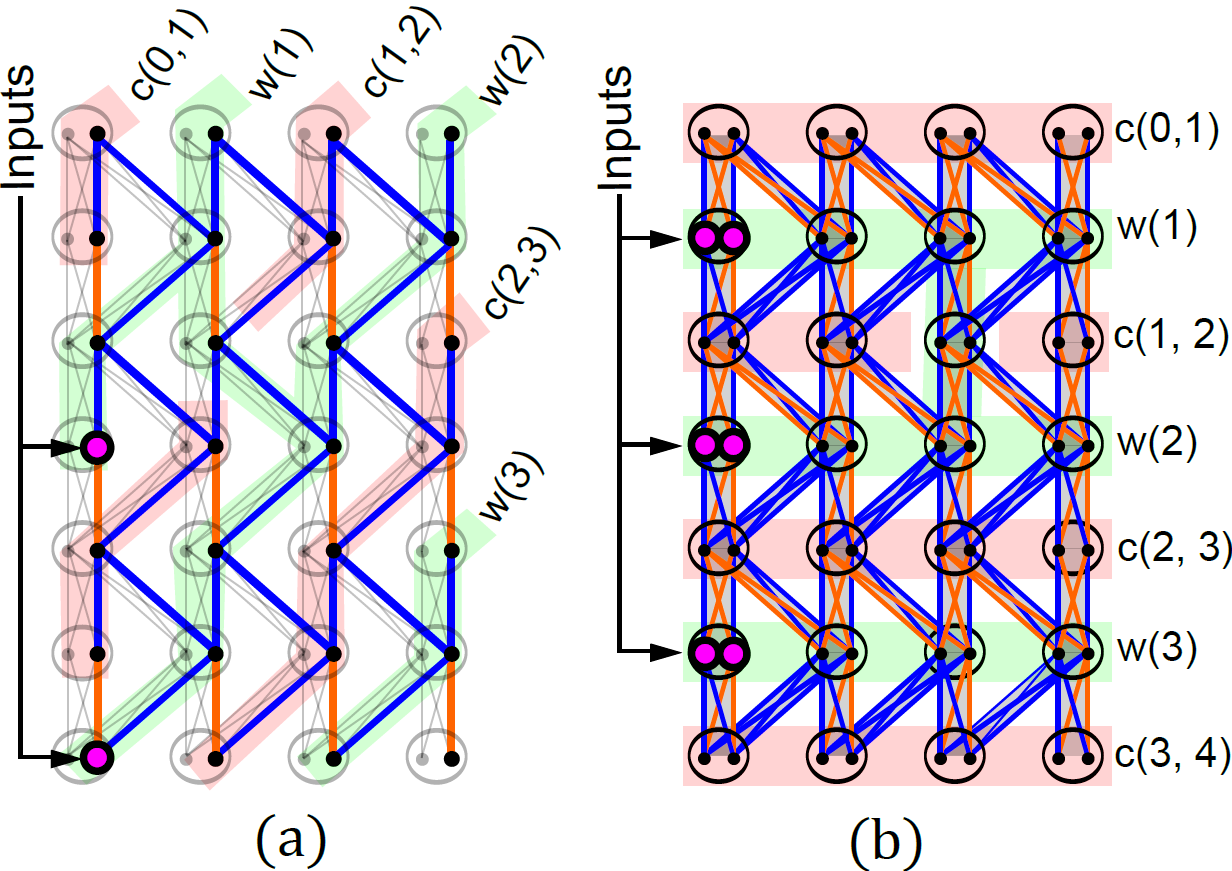}
\caption{Two MBQC protocols applied on the bilayer square-lattice cluster state $(\mathcal C = 2^{-3/2})$. In both cases, information is encoded on the left (in purple nodes) and flows from left to right along (green) wire segments. Wires are separated by lines of sacrificial qumodes (shown in the red segments).  These are referred to as \emph{control macronodes}~$\text{c}(i, i+1)$ because measurements on them control whether one- or two-qumode gates are applied on the adjacent \emph{wire macronodes}~$\text{w}(i)$ and~$\text{w}(i+1)$.  Two-qumode gates are represented by a connecting green segment between two adjacent wires. \textbf{(a)}~Standard one-way protocol~\cite{Menicucci2006,Gu2009} applied to the BSL graph after measuring $\hat q$ on all $Y$-polarized qumodes (shown faded). Time-ordering of the nodes has been preserved, resulting in an atypical nodal arrangement of an ordinary square-lattice graph. Information propagates at $45^{\circ}$ to the direction of increasing temporal index. Control nodes are measured in the $\hat{q}$-basis to delete them or in a different basis to implement a two-qumode gate. \textbf{(b)}~New MBQC protocol taking advantage of the BSL structure. Both layers of the lattice are used simultaneously, and quantum information propagates in the direction of increased temporal index, i.e.,~horizontally on the figure%
. Control and wire  macronodes  are now at a constant frequency, as shown. See text for further details.}
\label{main:fig:basicidea}
\end{figure}

The BSL CV cluster state is easily shown to be universal for MBQC. Simply measure $\hat{q}$ on all qumodes of one (e.g., $Y$) polarization, resulting in a CV cluster state with an ordinary square-lattice graph, which can be used with standard CV MBQC protocols~\cite{Menicucci2006,Gu2009}. This is shown in Fig.~\ref{main:fig:basicidea}(a).

Using so-called \emph{deletion} measurements (as above) to simplify the graph structure of a CV cluster state is a standard way to prove universality of a given graph~\cite{Flammia2009,Menicucci2011a}, but it is  a wasteful procedure to follow in practice  since half of the graph nodes and their connectivity are lost. More precisely, this method inefficiently uses available squeezing (and therefore entanglement~\cite{Braunstein2005}), which leads to extra noise when using these resources for quantum computing~\cite{Alexander2014}. Furthermore, lattice edges are at $45^{\circ}$ to the direction of increasing temporal index, meaning that either the information must flow in a zig-zag pattern or the lattice width will have to scale linearly with the length of the computation, hindering the scheme's scalability.

Fortunately, there exists a more favorable MBQC protocol that eschews all these complications and makes better use of the structure of the BSL CV cluster state, while still using just single-site measurements. The idea is to use both layers of the graph simultaneously and in a way analogous to the conventional (single-layer) square lattice protocol, as shown in Fig.~\ref{main:fig:basicidea}(b). Each lattice site, which we call a \emph{macronode}~\cite{Flammia2009}, is composed of two qumodes (one of each polarization). Qumodes with even node index carry the quantum information to be processed and are therefore called  \emph{wire macronodes}  (for `quantum wires'). Those with odd node index control the connectivity between the wires and are therefore called  \emph{control macronodes}.  Input states are localized with respect to the macronode structure and are encoded within the symmetric subspace of each  macronode (defined in the section below). One- and two-qumode Gaussian gates  are selected by the choice of homodyne measurement angles.

 To simplify the presentation, we will introduce our protocol within the context of an infinitely-squeezed BSL resource state. Any physical CV cluster state can only be finitely squeezed~\cite{Menicucci2006, Menicucci2011}, and this leads to introducing finite squeezing effects into the computation~\cite{Alexander2014}, which we discuss in Sec.~\ref{sec:noise}. 
 
\subsection{Computing with macronodes}
\label{subsec:computemacro}

%

For a given macronode with node index~$m$, comprised of individual qumodes labeled $Y$ and $Z$, we define the symmetric~($+$) and anti-symmetric~($-$) qumodes via 
\begin{align}
\label{main:eq:mapphystolog}
	\hat a_{m\pm} \coloneqq \frac {1} {\sqrt 2} (\hat a_{mZ} \pm \hat a_{mY})\,,
\end{align}
which is mathematically equivalent to a $\tfrac \pi 4$ polarization rotation into the diagonal and anti-diagonal qumode decomposition (equivalently, a 50:50 beamsplitter between the two qumodes).  Input states at a particular time step will either be the output state from the previous time step or new states directly injected into the BSL via an optical switch~\cite{Yokoyama2013}. They are localized to macronodes but distributed (symmetrically) between the two physical qumodes. 
We further define quadrature operators $\hat q$ (position) and~$\hat p$ (momentum) for each qumode through $\hat a = \tfrac {1} {\sqrt 2} (\hat q + i\hat p)$,  which satisfy $[\op q, \op p] = i$ with $\hbar = 1$.

Before describing our measurement protocol, we also provide some definitions for useful CV logic gates. These include an optical phase delay by~$\theta$,
\begin{align}
\op R(\theta) &\coloneqq \exp(i \theta \op a^\dag \op a) %
	= \exp{\left[ \frac{i \theta}{2} (\hat{q}^{2}+ \hat{p}^{2}-1)\right]}\,,
\end{align}
and a 50:50 beamsplitter between qumodes~$i$ and~$j$,
\begin{align}
\label{eq:BS5050}
	\op B_{ij} &\coloneqq \exp \left[-\frac \pi 4 (\op a_i^\dag \op a_j - \op a_j^\dag \op a_i) \right] \nonumber \\
	&= \exp\left[-i\frac \pi 4(\hat{q}_{i}\hat{p}_{j}-\hat{q}_{j}\hat{p}_{i})\right]\,.
\end{align}
We also define a (nonstandard) single-qumode squeezing operation:
\begin{align}
\op S(s) &\coloneqq \op R(\Im \ln s) \exp\left[-\frac 1 2 (\Re \ln s) (\op a^2-\op a^{\dag2}) \right] \nonumber \\
	&= \op R(\Im \ln s) \exp\left[-\frac i 2 (\Re \ln s) (\hat{q}\hat{p}+\hat{p}\hat{q}) \right]\,, 
\end{align}
where $s$ is known as the \emph{squeezing factor}. This gate is just an ordinary squeezing gate with \emph{squeezing parameter} ${r=\ln \abs{s}}$, followed by a $\pi$ phase delay if and only if ${s<0}$. We chose this form of the gate so that for all real $s \neq 0$, its Heisenberg action on the quadratures is $\op S(s)^\dag \op q \op S(s) = s\op q$ and $\op S(s)^\dag \op p \op S(s) = s^{-1}\op p$.

As is standard in MBQC, once the entangled resource state is prepared, quantum computation proceeds solely through adaptive local measurements.
Here we restrict the measurements to linear combinations of the quadrature operators, which will be shown to be sufficient to implement arbitrary multi-qumode Gaussian unitaries. Experimentally, this can be performed through homodyne detection. For any given qumode, we define the rotated quadrature operators%
\begin{align}
\hat{\mathbf{x}}(\theta) =  \begin{pmatrix} \hat{q}(\theta) \\ \hat{p}(\theta) \end{pmatrix} &\coloneqq  \op R^{\dagger} (\theta)\begin{pmatrix}  \hat{q} \\ \hat{p} \end{pmatrix}\op R(\theta)    \nonumber \\  & = \begin{pmatrix} \cos{\theta} & -\sin{\theta} \\  \sin{\theta} & \cos{\theta} \end{pmatrix} \begin{pmatrix}  \hat{q} \\ \hat{p} \end{pmatrix},
\label{main:eq:homodyne}
\end{align}
where the second line shows the symplectic-matrix representation of the Heisenberg action of a phase delay by $\theta$~\cite{Alexander2014}.

\begin{figure*}[t!]
\includegraphics[width=2\columnwidth]{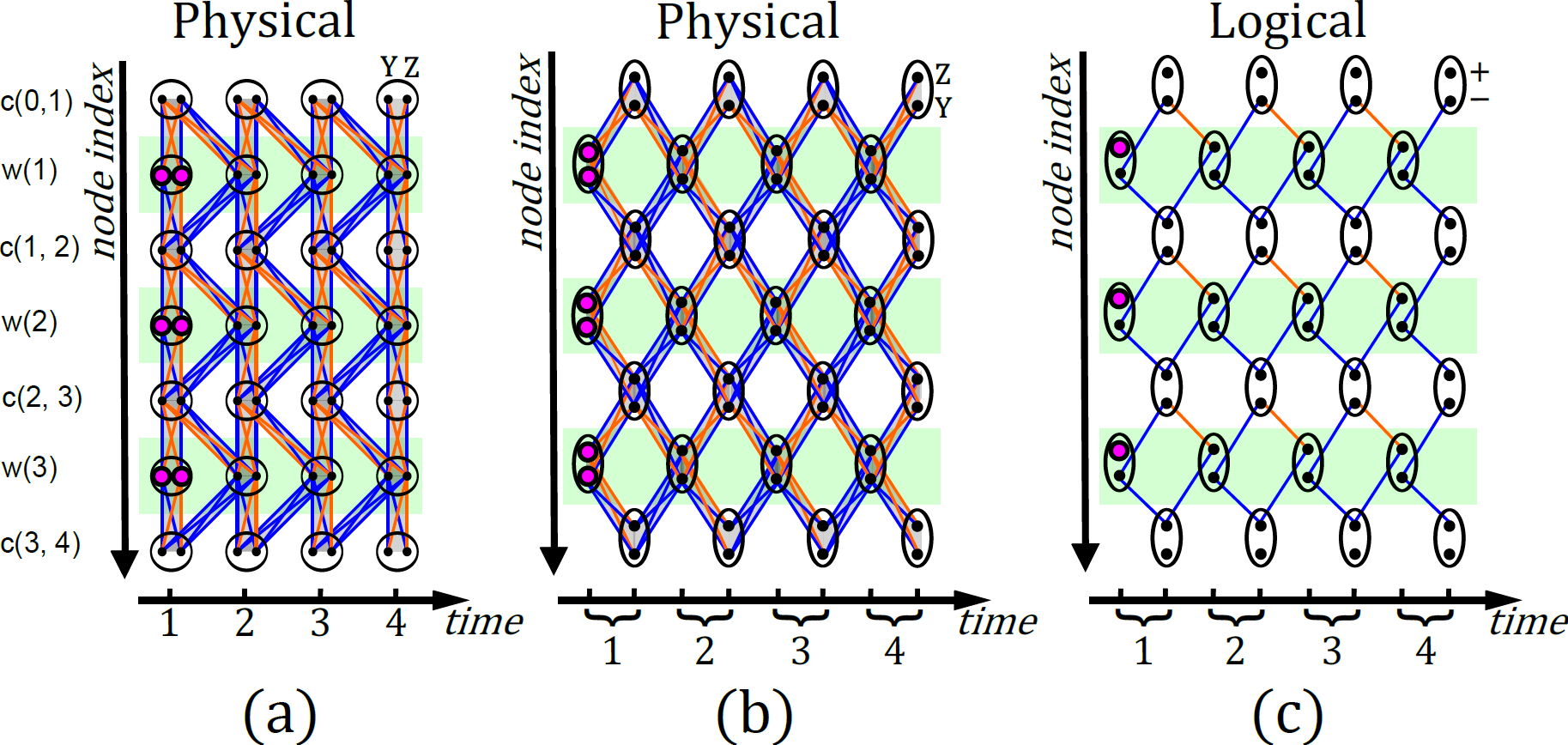}
\caption{ \textbf{(a)}~Simplified graphical-calculus representation~\cite{Menicucci2011a} of the bilayer square-lattice (BSL) CV cluster state.  Here, qumodes are ordered according to temporal index. Input states are encoded within macronodes on the left, shown in purple. Here and also in (b), $\mathcal{C} = (2\sqrt{2})^{-1}$. \textbf{(b)} As in (a) but time ordering has been partially sacrificed in order to make the square-lattice graph structure more apparent. \textbf{(c)} Each macronode is now represented in terms of the \emph{logical-mode} tensor-product structure [see Eq.~\ref{main:eq:mapphystolog}]. We use the same time-ordered node arrangement is as in (b). Unlike in the previous subfigures, here the graph has a lower connectivity [it is a disjoint collection of square graphs] and all input states are localized. In this subfigure,  $\mathcal C = 2^{-1/2}$.  We indicate internal qumode labeling on the top right macronode of each lattice.}
\label{fig:altrep}
\end{figure*}


  In \fig{altrep} we show alternative (and equivalent) graphical representations of the BSL CV cluster state.  Recall that within each macronode the map from the physical and logical mode labels is given by Eq.~\ref{main:eq:mapphystolog}. We can apply this map to every macronode, giving us a graph where each node now represents the \emph{symmetric} or \emph{anti-symmetric} mode of the enclosing macronode, as shown in \fig{altrep}(c). 
This graph reveals a simpler underlying logical structure that will provide us with a convenient framework for describing how homodyne measurements on the physical modes can implement useful gates.

Generically, due to the non-local nature of the map from physical ($Z,Y$) to logical ($+,-$) mode labels, local measurements on the physical modes will effectively ``stitch together" the disjoint square graphs present in \fig{altrep}(c).  
For a macronode $m$, measurement of {$\hat{p}_{mZ}(\theta_{mZ})$ and $\hat{p}_{mY}(\theta_{mY})$} can be represented by the following quantum circuit: 
\begin{align}
\includegraphics[width=0.8\columnwidth]{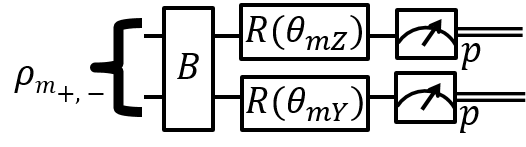}, \label{meascirc1}
\end{align}
where $\rho_{m_{+,-}}$ denotes the input state with respect to the logical ${(+, -)}$ mode tensor product structure. Above, the 50:50 beamsplitter takes us from logical to physical mode labels.  Rotated quadratures are measured as in Eq.~\ref{main:eq:homodyne}.  Such measurements on the wire macronodes connect square subgraphs  with their neighbors in the horizontal direction,  enabling ``wire-like" transmission along the BSL. The measurements on the control macronodes connect these neighboring wires vertically.

\subsection{Keeping square graphs disconnected}
\label{subsec:donotconnect}

For a fixed macronode,  there exists a one-parameter class of homodyne angles that do not connect the adjacent  square  graphs. 
Specifically, 
when ${\theta_{mZ}=\theta_{mY}=\theta}$, the above circuit (\ref{meascirc1}) is equivalent to 
\begin{align}
\includegraphics[width=0.8\columnwidth]{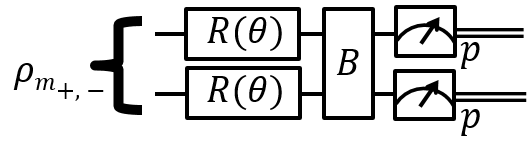}, \label{meascirc2}
\end{align}
where the single-qumode rotation gates  commute with the 50:50 beamsplitter because the rotation angles are the same~\cite{Gabay2016}.  This in turn is equivalent to
\begin{align}
\includegraphics[width=0.8\columnwidth]{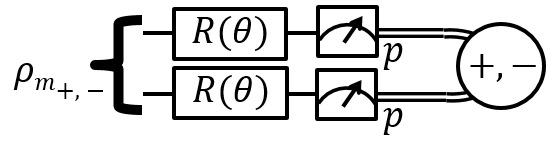}, \label{meascirc3}
\end{align}
where all circuit elements are now local, and we take the sum and difference of the measurement outcomes.  Therefore, choosing $\theta_{mZ}=\theta_{mY}$ for a particular wire or control macronode $m$ will disconnect the neighboring regions of the BSL graph in the horizontal or vertical direction, respectively. By restricting \emph{all} control macronode measurements in this way and including the required post processing (i.e., sum and difference of outcomes), the disjoint square graphs of \fig{altrep}(c) remain uncoupled by homodyne measurements with respect to the physical modes.
%
%

\begin{figure}[t!]
\includegraphics[width=1\linewidth]{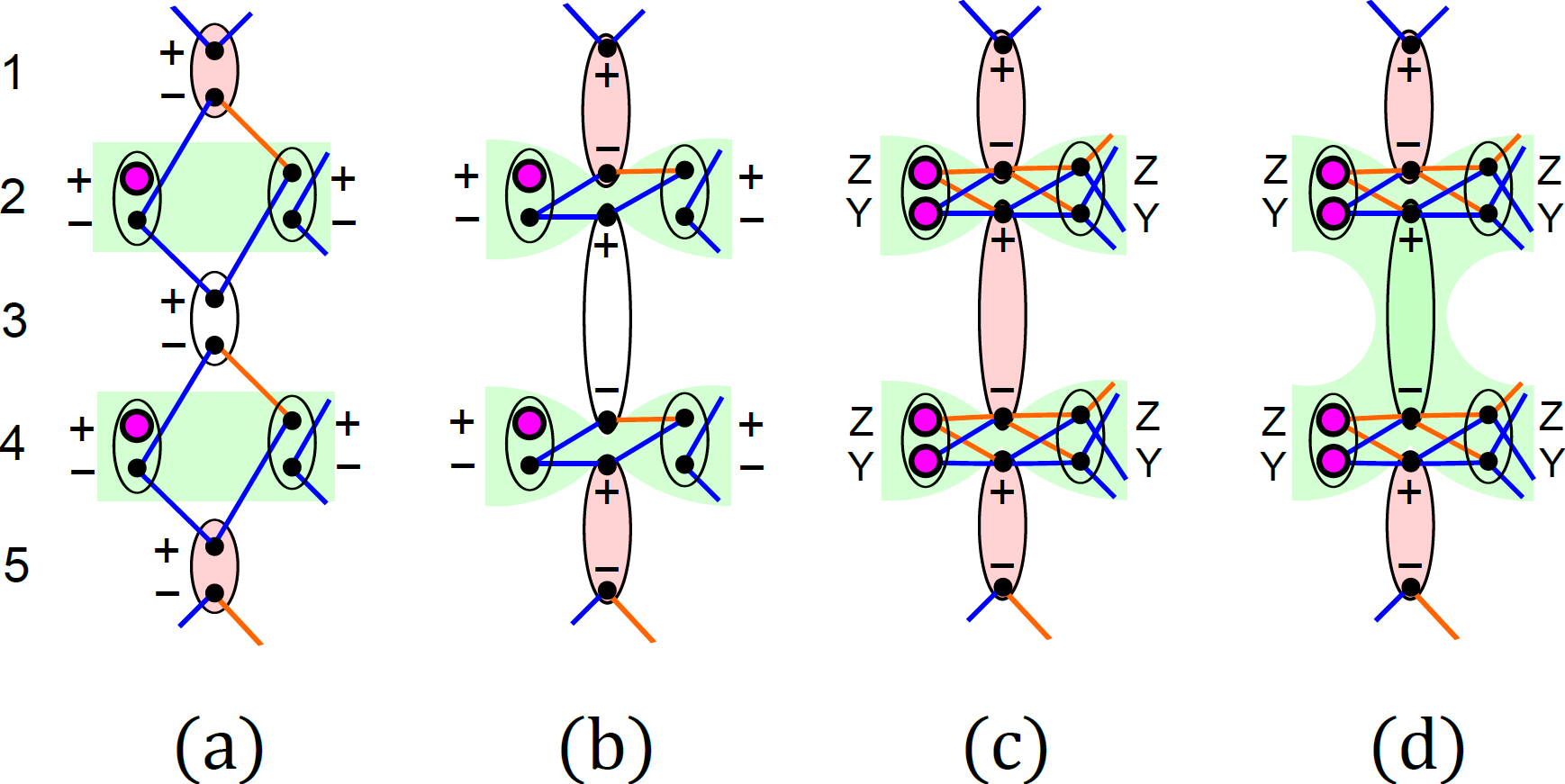}
\caption{(Color online) Implementing single- and two-qumode gates on the bilayer square lattice. Node indices [Eq.~\eqref{main:eq:nodeindex}] for all macronodes are provided on the left. A red ellipse indicates a restriction on the measurements of that macronode---specifically, $\theta_{cZ}=\theta_{cY}$ for control macronode~$c$. \textbf{(a)}~We begin with the configuration as in \fig{altrep}(c), with each macronode decomposed into logical $(+, -)$ modes. Note that measurements on macronodes 1 and 5 are restricted (red coloring). This decouples the two fully displayed square graphs from their partially displayed neighbors above and below. \textbf{(b)}~Same as~(a), except we visually separate the internal nodes of the control macronodes. \textbf{(c)}~Starting with~(b), we decompose the wire macronodes (within the green regions) with into physical $(Z,Y)$ modes in order to reveal a pair of CV dual-rail wires~\cite{Alexander2014, Yokoyama2013, Chen2014}. Restricting the measurements (red ellipse) of control macronode~3 allows one to implement single-qumode gates~\cite{Alexander2014} on each wire independently [Sec.~\ref{sec:SMgates}].
 \textbf{(d)}~Alternatively, if we set $\theta_{3Z}\neq\theta_{3Y}$, then control macronode 3 will mediate an entangling gate between the two neighboring wires [Sec.~\ref{sec:tmgates}].
}
\label{fig:diamonduncouple}
\end{figure}

\section{Universal gate set}
\label{sec:univgateset}

The methods above allow us to apply single-qumode gates on adjacent wires without them interacting. Alternatively, relaxing the restriction on a particuar control macronode implements a two-qumode gate between the adjacent wires at that location. In this section, we elaborate on this and construct a universal gate set for quantum computation on the BSL.%
%
%

\subsection{Single-qumode gates}\label{sec:SMgates}

\fig{diamonduncouple}(a)--(c) shows a new way to represent the BSL such that all measurements are local, but with respect to a mixture of physical ($Z,Y$) and logical ($+,-$) mode labels.  As information propagates along the lattice in the direction of increasing time index, information will flow strictly in the horizontal direction, and there will be no interactions between neighboring wire macronodes. 

 The structure shown in \fig{diamonduncouple}(c) is  identical to a collection of  CV \emph{dual-rail quantum wires}~\cite{Menicucci2011a,Alexander2014},  which are resources for universal single-qumode quantum computation. Therefore,  we can implement single-qumode gates on the BSL by directly implementing the \emph{macronode protocol} for the CV dual-rail wire from Ref.~\cite{Alexander2014}. 
 We briefly review it here.   If the qumodes at the left-most wire site are measured in the bases $\op p_{mZ}(\theta_{mZ})$ and $\op p_{mY}(\theta_{mY})$, as depicted here,
\begin{align}
\includegraphics[width=0.75\columnwidth]{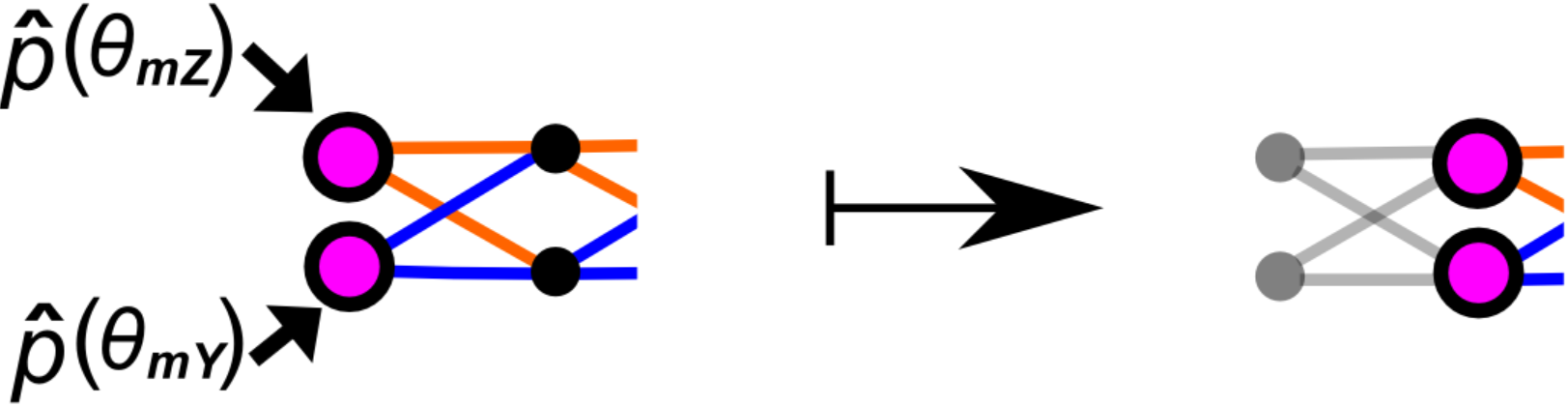}\,,
\end{align}
then (up to a displacements conditioned on the measurement outcomes and neglecting the effects of finite squeezing---see Sec.~\ref{sec:noise}) the overall Gaussian unitary applied to the encoded input state is~\cite{Alexander2014} $\op V(\theta_{mZ}, \theta_{mY})$, where
\begin{align}
	\op V(\theta_j, \theta_k) = \op R(\theta_{+})\op S(\tan\theta_{-})\op R(\theta_{+})
\label{eq:SMgate2}
\end{align}
with ${\theta_{\pm}=\tfrac 1 2 (\theta_j\pm\theta_k)}$. This is the basic building block for all single- and two-qumode gates that can be implemented on the BSL.

There is an important difference between the conventional CV dual-rail wire and the BSL, however. With respect to the original BSL time-ordered node layout [see \fig{altrep}(a)], it is natural to consider a single measurement step as translating input states horizontally by one time step---from wire macronode to wire macronode. This corresponds to translating two sites along the CV dual-rail wire since wire macronodes are interleaved by control macronodes,  as shown here:
\begin{align}
\includegraphics[width=0.7\columnwidth]{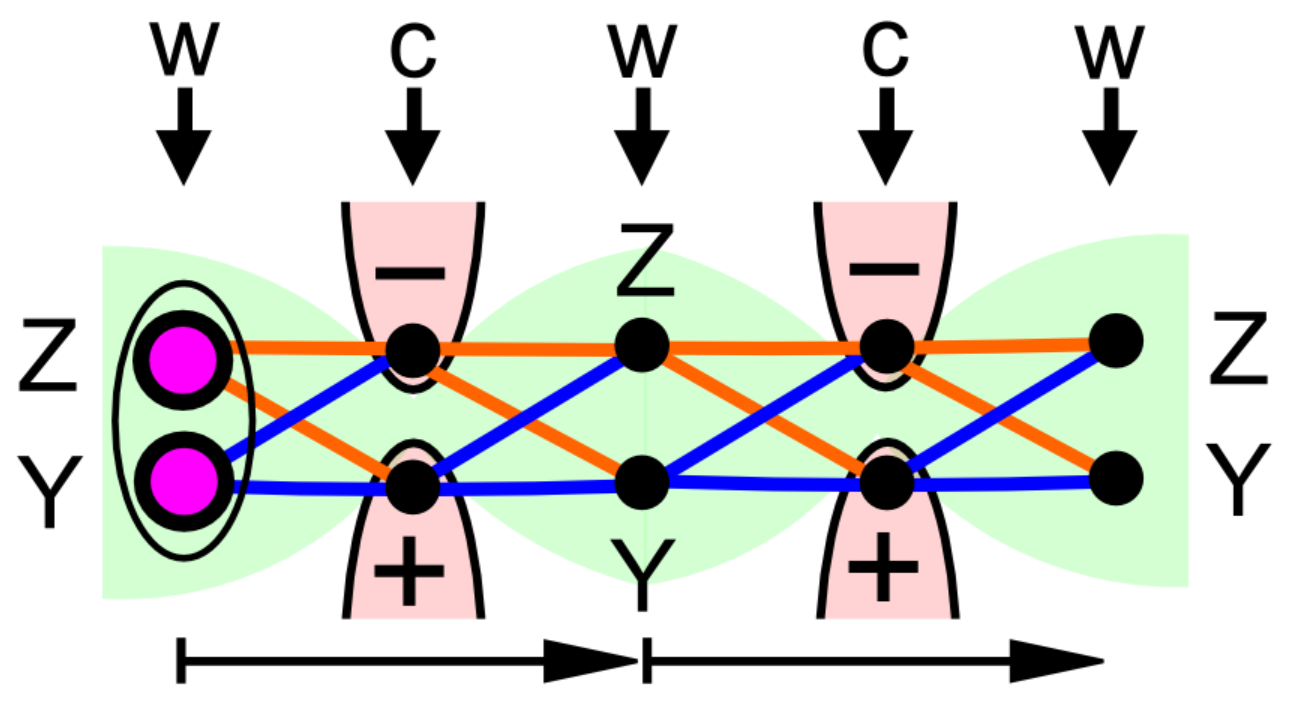}\,,
\end{align}
where the horizontal  black arrows indicate the size of each measurement step. 
Each measurement step implements two $\op V$~gates [Eq.~\eqref{eq:SMgate2}], but with one important caveat: the measurements on the control macronodes have to be constrained by the condition $\theta_{cZ}=\theta_{cY}$ (for control macronode~$c$) so that that neighboring wires decouple [as in Fig.~\ref{fig:diamonduncouple}(c)]. Note that these constraints jointly affect nodes of  separate neighboring wires, which share a control macronode.

Although these constraints do not completely specify the set of possible measurements on the control macronodes, some care has to be taken in assigning the measurement angles. For one thing, constraining all control macronode measurements to the same angle would projectively measure the encoded information (as discussed below in Sec.~\ref{subsec:projmeas}), thereby ending the computation at that point. On the other hand, %
attempting to use control macronode degrees of freedom to locally implement some desired gate on a particular wire would necessarily implement a nontrivial gate on both neighboring wires. For this reason, we fix all measurements on the control macronodes and only use the measurements on the wire macronodes to implement gates.

%

%

%

%

\begin{figure}
\includegraphics[width=0.9\columnwidth]{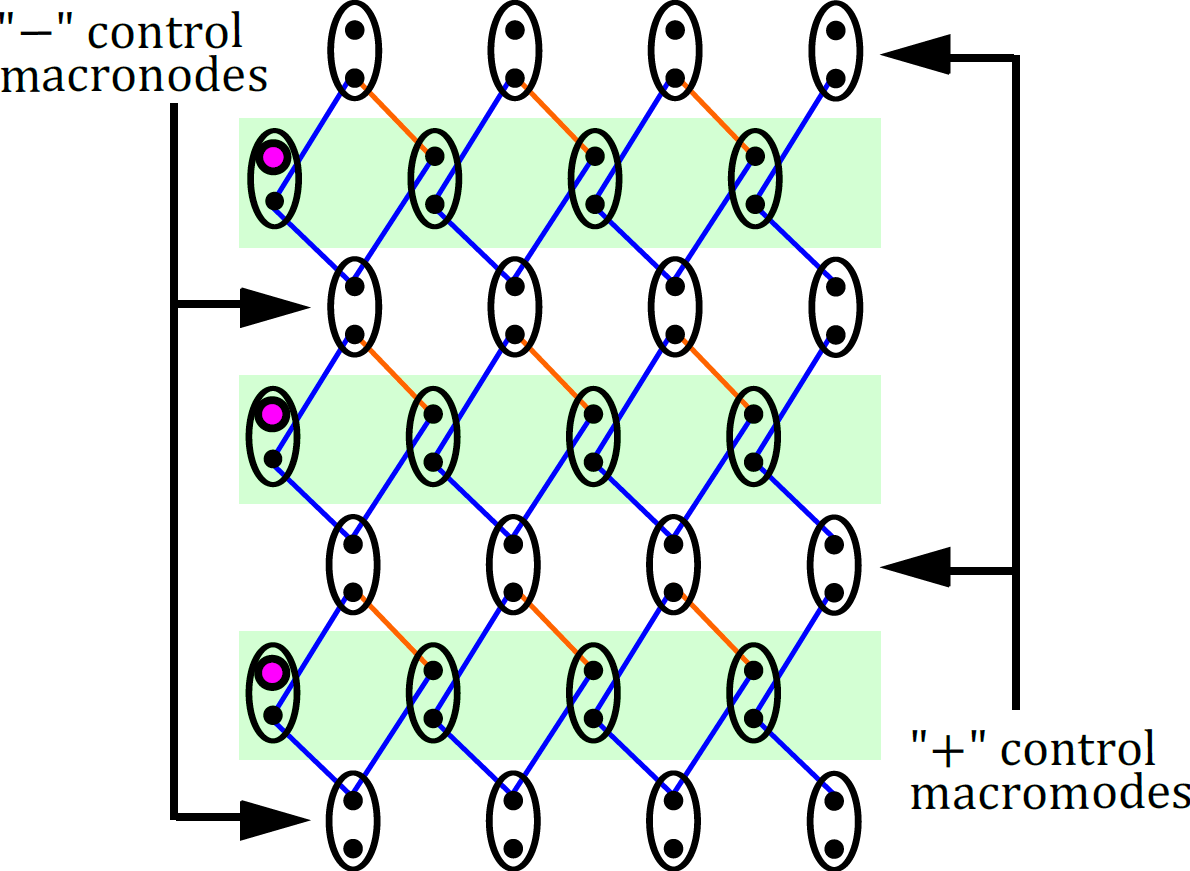}
\caption{(Color online) Sign convention for measurements on the control macronodes. 
Both physical modes of each $\pm$ control macronode (alternating top to bottom, as shown above) are measured in the basis specified by homodyne angle $\theta=\pm\frac{\pi}{4}$, respectively. }\label{fig:signconvention}
\end{figure}

A particularly convenient choice is to set the homodyne angles to be $\theta_{ c Z}={\theta_{ c Y}=\pm\frac{\pi}{4}}$ (for control macronode~$c$), where the sign alternates vertically with each row of control macronodes,
as shown in \fig{signconvention}. For one physical time step (i.e.,~measuring one wire macronode~$w$ and its neighboring control macronodes) on a  wire above  a row of $\pm$ control macronodes, this implements 
\begin{align}
\op V\left(\mp \frac \pi 4, \pm \frac \pi 4 \right)\op V(\theta_{wZ}, \theta_{wY}),
\end{align}
where the first gate $\op V(\theta_{wZ}, \theta_{wY})$ results from measurement of the wire macronode~$w$, and the second gate~$\op V(\mp \frac \pi 4, \pm \frac \pi 4 )$ results from the measurements of the two control macronodes above and below, as in Fig.~\ref{fig:altrep}(a). 
Plugging into Eq.~\eqref{eq:SMgate2}, we get %
\begin{align}
\op V\left(\mp \frac \pi 4, \pm \frac \pi 4 \right)&= \op S(\mp 1) %
\end{align}
Noting that $\op S(-1)\op V(\theta_{wZ}, \theta_{wY})=\op V(\theta_{wY}, \theta_{wZ})$,  
a single measurement step on the BSL implements 
$\op V(\theta_{wY}, \theta_{wZ})$ or $\op V(\theta_{wZ}, \theta_{wY})$, 
depending on whether the control macronodes below the wire are $+$ or $-$, respectively. 
Two applications of these gates generate all single-qumode Gaussian unitaries (up to displacements)~\cite{Yokoyama2013, Alexander2014}.

As mentioned above, we have neglected a ubiquitous phase-space displacement (dependent on the measurement outcomes) and the effects of finite squeezing in our discussion above. We did this in order to present clearly the basic logic of the protocol. The details of the additional displacements and squeezing effects can be found in Sec.~\ref{sec:noise}.

\subsection{Projective measurement}
\label{subsec:projmeas}

Notice that when ${\theta_{wZ} = \theta_{wY} = \theta}$, the squeezing term in Eq.~\eqref{eq:SMgate2} diverges. A gate is not applied in this case. Instead, this projectively measures both logical modes in $\op p(\theta)$, as can be seen from the symmetry discussed in Sec.~\ref{subsec:donotconnect}.

\subsection{Two-qumode gates}\label{sec:tmgates}

In the above, we found that (by appropriately restricting the control macronode measurements) we could treat the BSL as a collection of independent non-interacting quantum wires. This protocol can be extended to also include a two-qumode entangling gate by lifting the measurement restrictions on control macronodes that lie between neighboring wires. This corresponds to the case shown in \fig{diamonduncouple}(d). We parameterize the choice of measurements by the vector of homodyne angles ${\vec\theta = (\theta_{1Z}, \theta_{1Y}, \theta_{2Z},\dots \theta_{5Y})}$.

We would like our two-qumode gate protocol to be compatible with single-qumode gates applied on adjacent regions of the BSL. With respect to \fig{diamonduncouple}(d), we allow $\theta_{3Z}$ and $\theta_{3Y}$ to be free parameters, while $\theta_{1Z}=\theta_{1Y}=\theta_{5Z}=\theta_{5Y}=\pm\frac{\pi}{4}$, corresponding to macronode~3 being a $\mp$ control macronode, respectively. Correspondingly, selecting homodyne angles 
\begin{align}
	\vec\theta=\left(\pm\frac{\pi}{4}, \pm\frac{\pi}{4}, \theta_{2Z}, \theta_{2Y}, \theta_{3Z}, \theta_{3Y}, \theta_{4Z}, \theta_{4Y}, \pm\frac{\pi}{4}, \pm\frac{\pi}{4}\right)\label{eq:2modemeasangles}
\end{align}
implements a two-qumode gate whose form we will now derive.

\begin{figure*}[!t]
\includegraphics[width=1.9\columnwidth]{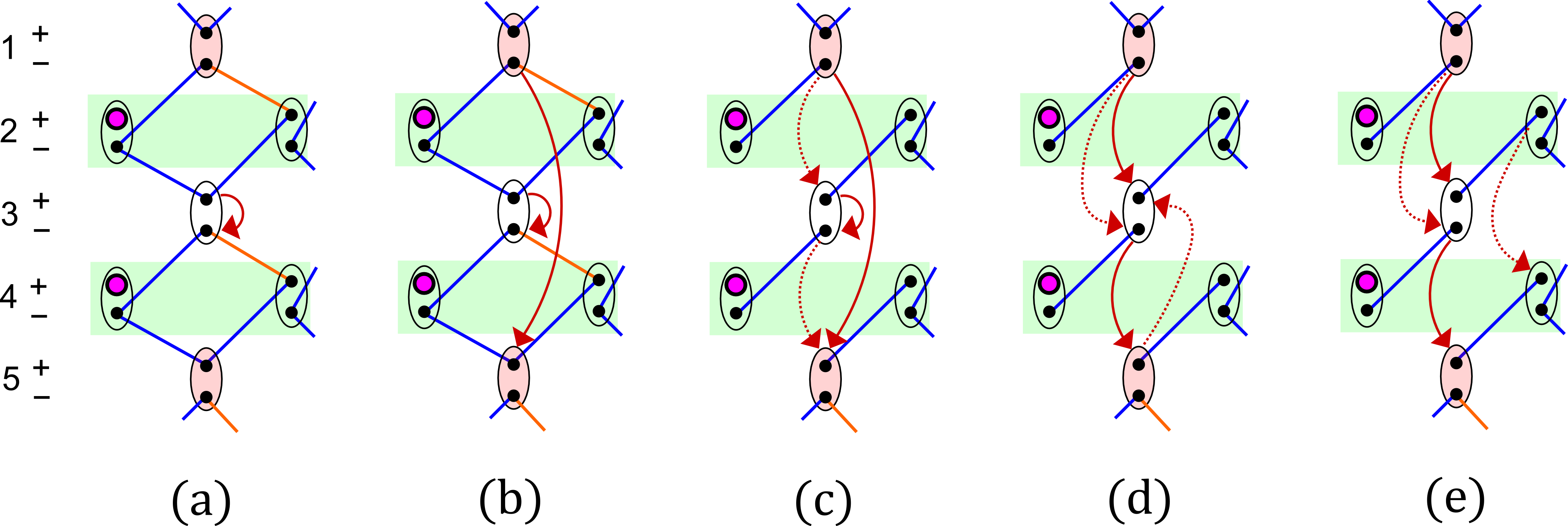}
\caption{(Color online) Beamsplitter gymnastics. All graphs are drawn in terms of logical~$(+,-)$ modes. A 50:50 beamsplitter~$\op B_{ij}$ between two qumodes~$i$ and~$j$ is indicated by a red arrow from~$i$ to~$j$. Where applicable, dashed-arrow beamsplitters always act before solid-arrow beamsplitters. \textbf{(a)}~We start from Fig.~\ref{fig:diamonduncouple}(a). Measuring control macronode~3 in the physical $(Z,Y)$ modes is equivalent to performing a beamsplitter as shown and then measuring in the logical modes.  \textbf{(b)}~Since all qumodes of control macronodes~1 and~5 are measured in the same basis, we are free to insert an additional beamsplitter between them as shown [see Sec.~\ref{subsec:donotconnect}]. This is the key observation.
 \textbf{(c)}~The squares in~(b) (with ${\mathcal C = 2^{-1/2}}$) can be replaced by pairs of two-qumode CV cluster states (with $\mathcal C=1$) followed by two additional beamsplitters as shown (dashed)~\cite{Menicucci2011a}. These occur before the other two (solid). \textbf{(d)}~By direct calculation using their symplectic representation~\cite{Menicucci2011}, $\op B_{il}\op B_{jk}(\op B_{kl}\op B_{ij}) = (\op B_{kl}\op B_{ij})\op B_{lj}\op B_{ik}$. \textbf{(e)}~The symmetries of a pair of two-qumode CV cluster states (see Appendix~\ref{sec:symmcvcs}) allow for the beamsplitter to be moved to the other two qumodes as shown.}
\label{fig:2modeproof}
\end{figure*}

Our strategy for the derivation will be to use symmetries of CV cluster states and ``beamsplitter gymnastics'' to reduce the evolution to a form that can be interpreted as a combination of two steps of evolution on the CV dual-rail wire~\cite{Alexander2014} interleaved with two additional beamsplitters. To this end, we call attention to Fig.~\ref{fig:2modeproof}, which shows that measurements on the original resource shown in~(a) are equivalent to the same measurements on the resource shown in~(e). As such, we can read off the evolution from the last subfigure, using knowledge of evolution on the CV dual-rail wire~\cite{Alexander2014}.

We summarise this procedure here, referring to Fig.~\ref{fig:2modeproof}(e). %
First, the leftmost wire macronodes (2~and~4) are measured, applying $\op V(\theta_{2Z}, \theta_{2Y})\otimes\op V(\theta_{4Z}, \theta_{4Y})$ to the input and teleporting the output into qumodes~$1-$ and~$3-$, respectively. Then, the 50:50 beamsplitter between those two qumodes (dotted arrow) is applied. %
Next, the solid-arrow beamsplitters and measurements of the control macronodes
implement the gate~$\op V(\pm\frac{\pi}{4}, \theta_{3Z})\otimes\op V(\theta_{3Y}, \pm\frac{\pi}{4})$, teleporting the output to qumodes~$2+$ and~$4+$ at the following timestep. Finally, the last dotted-arrow beamsplitter acts on this output, concluding the evolution. 

Thus, up to displacements and neglecting finite-squeezing-induced noise (see Sec.~\ref{sec:noise}), the total gate applied is the combination of all of these individual gates:
\begin{multline}
\op B_{2+, 4+} \left[\op V \left(\pm\frac{\pi}{4}, \theta_{3Z}\right) \otimes \op V \left(\theta_{3Y}, \pm\frac{\pi}{4}\right)\right] \\
\times \op B_{2+, 4+} \bigl[\op V \left(\theta_{2Z}, \theta_{2Y}\right)\otimes \op V \left(\theta_{4Z}, \theta_{4Y}\right) \bigr], 
\label{eq:gen2gate1}
\end{multline}
where the tensor product is $\mathcal H_{(2+)}\otimes \mathcal H_{(4+)}$. %
This captures the most general type of two-qumode Gaussian unitary gate compatible with our framework.

Though we have the general form, it is useful to give particular measurement parameters that reduce the two-qumode gates into a simple form. It is also desirable to choose a form that is commonly included in universal gate sets, such as the  CV  controlled-Z ($\CZ$) gate~\cite{Gu2009}, defined as  $\op C_{Z}(g)= \exp{\left[i g \hat{q}\otimes\hat{q}\right]}$. 

While there is no valid choice of measurement parameters in $\vec\theta$ that yields an exact $\CZ$ gate, it is possible to implement one followed by phase delays that can be corrected in the next step by applying the single-qumode measurement protocol immediately after this gate. Again assuming macronode~3 is a $\mp$ control macronode, choosing 
\begin{align}
\label{eq:thetaCZ}
	\vec\theta= %
	\pm\left(\frac{\pi}{4}, \frac{\pi}{4}, -\frac{\pi}{8}, \frac{3\pi}{8}, \frac \pi 4 \pm \phi, \frac \pi 4 \mp \phi, -\frac{\pi}{8}, \frac{3\pi}{8}, \frac{\pi}{4}, \frac{\pi}{4}\right)
\end{align}
reduces \eq{gen2gate1} to
\begin{align}
\label{eq:CZgate}
	\left[\op R\left( \mp \frac{3\pi}{4} \right)\otimes\op R\left(\pm\frac \pi 4\right)\right]
	\CZ(2\cot \phi)
	\,,
\end{align}
which is a tunable-strength $\CZ$ gate followed by (known, fixed) phase delays that can be undone at the next time step. Appendix~\ref{app:CZcalcs} contains the detailed derivation. Once again, we postpone discussing finite-squeezing effects and outcome-dependent displacements until Sec.~\ref{sec:noise}.

\subsection{Alternative representation of two-qumode gate implementation}\label{sec:altrepgc}

In the previous two subsections, 
we showed  how measurements on the control macronodes  selected between applying either a pair of single-qumode gates or a two-qumode gate  on neighboring wires. Here we provide an alternative description of this mechanism that employs more fully the graphical calculus for Gaussian pure states~\cite{Menicucci2011}.
 
\begin{figure}
\includegraphics[width=1\columnwidth]{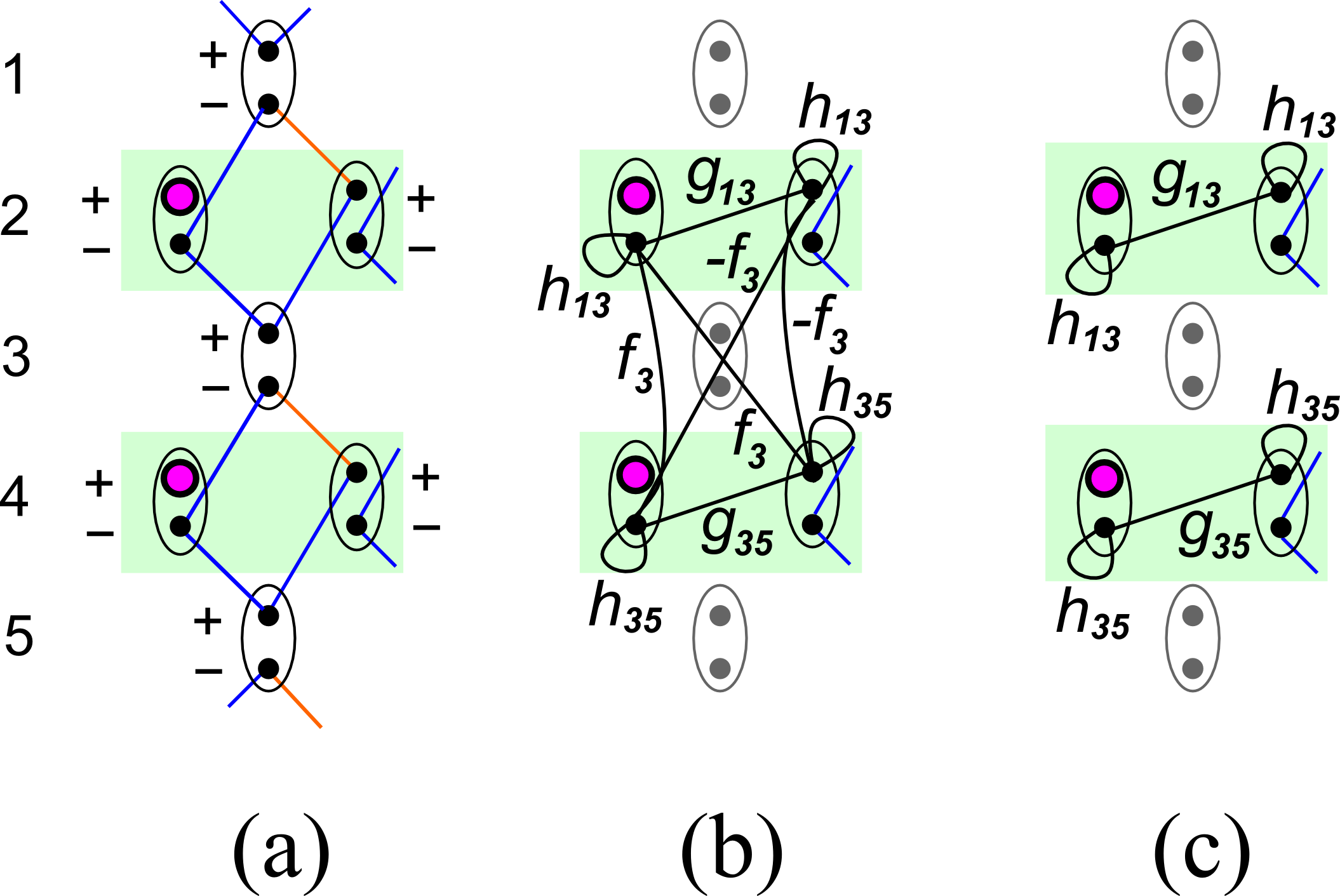}
\caption{%
Graphical-calculus representation~\cite{Menicucci2011} of measurements on a subregion of the bilayer square lattice with inputs in purple.  Here we show how measuring the control macronodes $(1,3,5)$ in two different ways leads to different connectivities of the wires.  
\textbf{(a)}~The lattice prior to measurement. We assume that $\theta_{1Z}=\theta_{1Y}$ and $\theta_{5Z}=\theta_{5Y}$. Having $\theta_{3Z}\neq\theta_{3Y}$ or $\theta_{3Z}=\theta_{3Y}$ will result in a graph as shown in subfigures (b) and (c), respectively. \textbf{(b)}~After measurement, the resulting graph has connecting edges between the wire macronodes. This is consistent with the application of a two-qumode gate between the encoded inputs as was shown in Sec.~\ref{sec:tmgates}.  Relevant graphical weights are defined in Eqs.~\eqref{eq:cweight}, \eqref{eq:eweight}, and \eqref{eq:gweight}.  \textbf{(c)}~After measurement, there are no graph edges connecting the input macronodes. Therefore, performing measurements on the input macronodes results in the application of single-qumode gates only. Thus, these entangled pairs can be thought of as separate quantum wires. Note that %
the four remaining edge weights share a dependence on $\theta_{3Z}$. In other words, the weights of adjacent quantum wires---and hence the single-qumode gate applied on them---are logically dependent in general. This is consistent with what was shown in Sec.~\ref{sec:SMgates}. Unlabeled edges all have  $\mathcal{C}=2^{-1/2}$. }
\label{fig:control2mode}
\end{figure}

Rather than finding a graphical description of the BSL that uses a mixture of physical $(Z, Y)$ and logical $(+,-)$ mode labels as in \fig{diamonduncouple}, we can instead consider the graphical representation of the premeasurement of the control macronodes (in analogy to ``wire shortening" in cluster state terminology~\cite{Gu2009}), as shown in \fig{control2mode}. Note that for the measurement-based implementation of Gaussian gates,  cluster nodes can be measured in any order  since the result is equivalent up to a final phase-space displacement~\cite{Gu2009}. 

The edge weights that are changed by the measurements are functions of the homodyne angles  on  the control macronodes and are given below in the large-squeezing limit. We get these from the graph transformation rules~\cite{Menicucci2011} corresponding to homodyne measurements on the physical modes of the BSL and then taking the limit  ${r\to\infty}$ (we choose this limit for clarity of presentation only). %
The edge weights in Fig.~\ref{fig:control2mode} are
\begin{align}
f_{i} &= \frac{1}{4} ( \cot{\theta_{iZ}}-\cot{\theta_{iY}}), \label{eq:cweight} \\
h_{ij} &= \frac{1}{4} ( -\cot{\theta_{iZ}}-\cot{\theta_{iY}} - \cot{\theta_{jZ}} - \cot{\theta_{jY}}),  \label{eq:eweight} \\
g_{ij} &=\frac{1}{4}( \cot{\theta_{iZ}}+\cot{\theta_{iY}}- \cot{\theta_{jZ}}-\cot{\theta_{jY}}).  \label{eq:gweight} 
\end{align} 

After the control macronodes are measured and when ${\theta_{3Z}\neq\theta_{3Y}}$, wire macronodes~2 and~4 are clearly connected by horizontal and diagonal links [see \fig{control2mode}(b)]. Attempting to ``teleport" the input states through this highly connected resource state will 
 entangle  
the input states.  Contrast this with the case when ${\theta_{3Z}=\theta_{3Y}}$ [see \fig{control2mode}(c)], where the resource state is simply a pair of unconnected entangled pairs.  The latter  is useful for propagating input states horizontally across the lattice without entangling the inputs~\cite{Alexander2014}.

%

\subsection{Achieving Universal Quantum Computation}

A CV controlled-$Z$ gate $\CZ$ can be applied between any two adjacent wires at any point on the BSL by locally substituting the macronode  protocol with the entangling-gate protocol described in Sec.~\ref{sec:tmgates}. This can be done repeatedly so long as each wire is involved in at most one $\CZ$ gate at a time. Together with vacuum input states and Weyl-Heisenberg displacement operations, these gates are universal for multimode Gaussian computation~\cite{Gu2009}.

In order to achieve universal quantum computation, we also need to include non-Gaussian resources~\cite{Gu2009}. (Sub-universal algorithms, such as state verification~\cite{Yokoyama2013,Chen2014}, only require homodyne detection.) In principle, it does not matter which type of resource is used~\cite{Lloyd1999}. Typical examples include photon-counting measurements supplemented with Gaussian resources to implement a cubic phase gate~\cite{Gottesman2001} or preparation and injection of non-Gaussian magic states such as photon subtracted states~\cite{Wenger2004}. We leave the detailed implementation to future work.

\section{Displacements and finite-squeezing effects}\label{sec:noise}

Thus far, we have neglected both the measurement-outcome-dependent displacements and finite squeezing effects that always arise in CV MBQC~\cite{Gu2009,Alexander2014}. We account for them here.

Since all evolution on the BSL can be reduced to evolution on the CV dual-rail wire (plus additional beamsplitters in the case of the two-qumode gate), all we need to do to take into account the effects of the measurement outcomes and finite squeezing is to use Eq.~(3.4) from Ref.~\cite{Alexander2014}, which amounts to 
replacing %
$\op V(\theta_j, \theta_k)$ [Eq.~\eqref{eq:SMgate2}] with
\begin{align}
\label{eq:noisyevol}
	\op V(r, m_j, m_k, \theta_j, \theta_k) %
	\coloneqq \op N(r) \op D(m_j, m_k, \theta_j, \theta_k) \op V(\theta_j, \theta_k),
\end{align}
where%
\begin{align}
\label{eq:dispop}
	\op D(m_j, m_k, \theta_j, \theta_k) = \op D\left[\frac{-i e^{i \theta_k} {m_j}-i e^{i \theta_j} {m_k}}{\sin (\theta_j-\theta_k)}\right]
\end{align}
is a phase-space displacement [${\op D(\alpha) = e^{\alpha \op a^\dag - \alpha^* \op a}}$] that depends on the homodyne angles and associated measurement outcomes~($m_j$, $m_k$), and
\begin{align}
\label{eq:noiseop}
	\op N(r) = e^{-\varepsilon \op q^2/2 } e^{-\varepsilon \op p^2/2t^2 }  \op S (t^{-1})
\end{align}
is a nonunitary operator that captures the effects of finite squeezing. We recover Eq.~\eqref{eq:SMgate2} in the limit of large squeezing and when all measurement outcomes are zero:
\begin{align}
	\op V(\theta_j, \theta_k) = \lim_{r\to \infty} \op V(r, 0, 0, \theta_j, \theta_k).
\end{align}
More generally, the displacements can either be actively corrected at each step or merely accounted for using feedforward~\cite{Gu2009}.

Noise from finite squeezing is ubiquitous in all MQBC protocols using CV cluster states, but fault tolerance is still possible using quantum error correction~\cite{Gottesman2001} provided that the overall squeezing levels---which set the amount of noise introduced per gate~\cite{Alexander2014}---are high enough~\cite{Menicucci2014}. The only known threshold result~\cite{Menicucci2014} states that no more than 20.5~dB of squeezing will be required. Squeezing levels in temporal-mode~\cite{Yokoyama2013} and frequency-mode~\cite{Chen2014} cluster-state experiments (5~dB and 3.2~dB, respectively) fall short of this, but state-of-the-art experiments in optics~\cite{Eberle2010} are within an order of magnitude (12.7~dB). The existence of a compact and scalable protocol such as the one presented here is likely to further spur on experimental and theoretical work to close this gap.

\emph{Technical note.}---The astute reader will note that this presentation differs from that of Ref.~\cite{Alexander2014} in three ways. First, the $r$-dependent squeezing term~$\op S(t^{-1})$ appears after the displacements in Eq.~\eqref{eq:noisyevol}, while it appears before them in Eq.~(3.4) of Ref.~\cite{Alexander2014}. We have modified our displacement operator~\eqref{eq:dispop} accordingly (Cf.~Eqs.~(3.8) and~(3.9) in Ref.~\cite{Alexander2014}), which allows us to group all finite-squeezing effects to the end and allows our displacement to depend only on the measurement angles and outcomes (and not on~$r$). Second, we have written the displacement in terms of the standard quantum-optics displacement operator, which relates to the Weyl-Heisenberg displacements as $\op X(s) \op Z(t) = (\text{phase}) \op D[(s+i t)/\sqrt 2]$, and we ignore the overall phase. Third, we have corrected a typo in Eqs.~(3.8) and~(3.9) in Ref.~\cite{Alexander2014}, which is that $\sin \theta_{i-}$ should actually be $\sin 2\theta_{i-}$.

\section{Conclusion}\label{sec:conc}

%
We have proposed an extremely compact and scalable method for producing---from a single OPO and simple interferometer---a continuous-variable~(CV) cluster state of unprecedented size [$(3\times10^3) \times \infty$] that is universal for quantum computation. The proposal has all the advantages of record-breaking temporal- and frequency-multiplexed schemes~\cite{Yokoyama2013,Chen2014} while vastly increasing the size of the lattice by utilising both types of multiplexing at once. This is the most compact and scalable proposal for CV cluster states to date, and it is implementable today using demonstrated quantum-optical technology.  In addition, we have generalized the one-way model for quantum computing to utilize the generated resource for quantum computation. The result translates familiar notions of CV measurement-based quantum computing~(MBQC) to the particular state proposed here, generalizing prior work based on one-dimensional, macronode-based CV cluster states%
~\cite{Yokoyama2013, Alexander2014}.

The vast majority of the existing literature on CV cluster states to date has treated canonical CV cluster states (i.e., those described in Refs.~\cite{Zhang2006,Menicucci2006,Gu2009}) as the appropriate target for an MBQC resource state. 
The work presented here---as well as the entire research direction upon which it is based---shows  that we should shift the focus onto CV cluster states with a macronode structure~\cite{Menicucci2008,Flammia2009,Menicucci2011a,Wang2014,Alexander2014,Yokoyama2013,Chen2014}. These schemes, which are all based on bipartite, self-inverse graphs~\cite{Flammia2009}, have been demonstrated to have unprecedented scalability~\cite{Yokoyama2013,Chen2014} and to admit novel, flexible~\cite{Alexander2016}, and more efficient~\cite{Alexander2014} quantum-computing schemes within the MBQC paradigm.

The work presented here further underscores this point, emphasizing the importance of bipartite, self-inverse graphs and of focussing on scalable designs from the ground up when working with CV cluster states.
One might hope that the optimized protocols available for these states~\cite{Alexander2014, Alexander2016}
could be used to improve the fault-tolerance threshold for MBQC using CV cluster states~\cite{Menicucci2014}. We leave this question to future work.

\emph{Note added in proof.} Recently, the state of the art in cluster state generation and optical squeezing has substantially improved. Temporal-mode one-dimensional CV cluster states have recently reached sizes of one million modes with no sign of phase drift over the course of the experiment~\cite{Yoshikawa2016}. Optical squeezing levels have also improved, with the state of the art now at 15 dB~\cite{Vahlbruch2016}.

\acknowledgments

This work was supported by the U.S. National Science Foundation under grant No.\ PHY-1206029 (for P.W., N.S., and M.C.'s graduate stipends and O.P.'s summer salary) and by the DARPA Quiness program (for O.P.'s summer salary).  R.N.A.\ and N.C.M.\ were  supported by the Australian Research Council under grant No.~DE120102204.

\appendix

\section{Graphical calculus for Gaussian pure states}

%
\label{sec:introGC}
  Any $N$-qumode Gaussian pure state $\ket{\psi_{\mathbf{Z}}}$ can be represented uniquely (up to phase-space displacement and overall phase) by an $N$-node, complex-weighted, undirected graph~\cite{Menicucci2011}.  This graph~$\mat Z$ can be represented pictorially or, equivalently, by a corresponding $N\times N$ complex-valued adjacency matrix
\begin{equation}
\mathbf{Z}\coloneqq\mathbf{V} + i \mathbf{U}, \label{eq:zmat}
\end{equation}
where $\mathbf{V}$ and $\mathbf{U}$ are $N\times N$ symmetric real-valued matrices, and $\mathbf{U} > 0$. This object is related to the wavefunction in the following way:
\begin{equation}
\psi_{\mathbf{Z}} (\mathbf{q}) =\frac{(\det{\mathbf{U}})^{1/4}}{\pi^{N/4}} \exp\left[ \frac{i}{2}\mathbf{q}^{T}\mathbf{Z}\mathbf{q}\right], \label{eq:graphwave}
\end{equation}

A covariance matrix for this state can be expressed in terms of the matrices in \eq{zmat}.  First, denote the vector of $2n$ position and momentum quadrature operators as 
 \begin{align}\mathbf{\hat{x}}\coloneqq (\hat{q}_{1}, \hat{q}_{2}, \dotsm \hat{q}_{N}, \hat{p}_{1}, \hat{p}_{2}, \dotsm \hat{p}_{N} )^{\text{T}}.
 \end{align}
 Then~\cite{Menicucci2011},
\begin{align}
\mathbf{\Sigma}\coloneqq \frac 1 2 \avg{\{\opvec x, \opvec x^\tp\}} = \frac{1}{2}\begin{pmatrix} \mathbf{U}^{-1} &  \mathbf{U}^{-1} \mathbf{V} \\ \mathbf{V} \mathbf{U}^{-1} & \mathbf{U} + \mathbf{V} \mathbf{U}^{-1}  \mathbf{V} \end{pmatrix}, 
\end{align}
 which in turn allows us to give an expression for the Wigner function:
\begin{align}
W(\mathbf{x} ) = (2 \pi )^{-N} (\det{\mathbf{\Sigma}})^{-1/2}\exp{\left[ -\frac{1}{2}\mathbf{x}^{\text{T}}\mathbf{\Sigma}^{-1}\mathbf{x}\right] }.
\end{align}
For some Gaussian unitary $\op U$, we can define $\ket{\psi_{\mathbf{Z^{\prime}}}}$ to be
\begin{equation}
\ket{\psi_{\mathbf{Z^{\prime}}}} \coloneqq \op U\ket{\psi_{\mathbf{Z}}}.
\end{equation}
 $\ket{\psi_{\mathbf{Z^{\prime}}}}$ is also a Gaussian pure state (by the definition of a Gaussian unitary).  How is the graph $\mathbf{Z}^{\prime}$ (corresponding to state $\ket{\psi_{\mathbf{Z^{\prime}}}}$) related to the original graph~$\mathbf{Z}$ by the Gaussian unitary? The Heisenberg action of $\op U$ on $\mathbf{\hat{x}}$  is linear, which means it can be represented as~\cite{Menicucci2011}
\begin{align}
\op U^{\dagger}\mathbf{\hat{x}}\op U=: \mathbf{S}_{\op U} \mathbf{\hat{x}}\,
\end{align}
where $\mathbf{S}_{\op U}$ is a ${2N\times 2N}$ symplectic matrix.  If we represent $\mathbf{S}_{\op U}$ as
\begin{equation} \label{eq:sdecomp}
\mathbf{S}_{\op U}=\begin{pmatrix} \mathbf{A} & \mathbf{B} \\ \mathbf{C} & \mathbf{D} \end{pmatrix},
\end{equation}
then the corresponding graph update rule is~\cite{Menicucci2011}
\begin{equation} \label{eq:graphtrans}
\mathbf{Z}\rightarrow \mathbf{Z}^{\prime}=(\mathbf{C}+\mathbf{D}\mathbf{Z} )(\mathbf{A} +\mathbf{B}\mathbf{Z} )^{-1}.
\end{equation}

\section{Simplified graphical calculus}

In general, representing all the features of $\mathbf{Z}$ requires an appropriately connected graph with all edges (including self-loops) labeled by complex-valued weights~\cite{Menicucci2011}. When representing Gaussian pure states with uniformly weighted graphs, it is convenient to employ a simplified set of rules. In the main text and wherever possible in the supplementary material, we represent Gaussian pure states using simplified graphs, as  introduced  in Ref.~\cite{Menicucci2011a}.  This allows us to represent graph edge weights by color and omit self-loops from the illustrations. 

With the exception of the (omitted) self-loop weights, the edge weights are implicitly defined as $\pm\mathcal{C} t$, where $\mathcal{C}$ is called the \emph{edge-weight coefficient} and can be thought of as the edge weight magnitude in the infinite squeezing limit, while 
\begin{align}
	t:=\tanh{2r}
\label{eq:tanh}
\end{align}
 can be thought of as a rescaling factor that depends on an overall squeezing parameter~$r$ for the state. In the graphs, signs of $+$ and $-$ are represented by blue and orange coloring, respectively, and $\mathcal{C}$ is indicated within relevant figure captions. Note that in the infinite squeezing limit (${r\rightarrow\infty}$), edge weight $\pm\mathcal{C}t \to \pm\mathcal{C}$. For all graphs, all black nodes have self-loop edges with weight $i \varepsilon$, where 
 \begin{align}
 \varepsilon \coloneqq \sech{2r}.\label{eq:epsilon}
 \end{align}

\begin{figure}
\includegraphics[width=0.8\columnwidth]{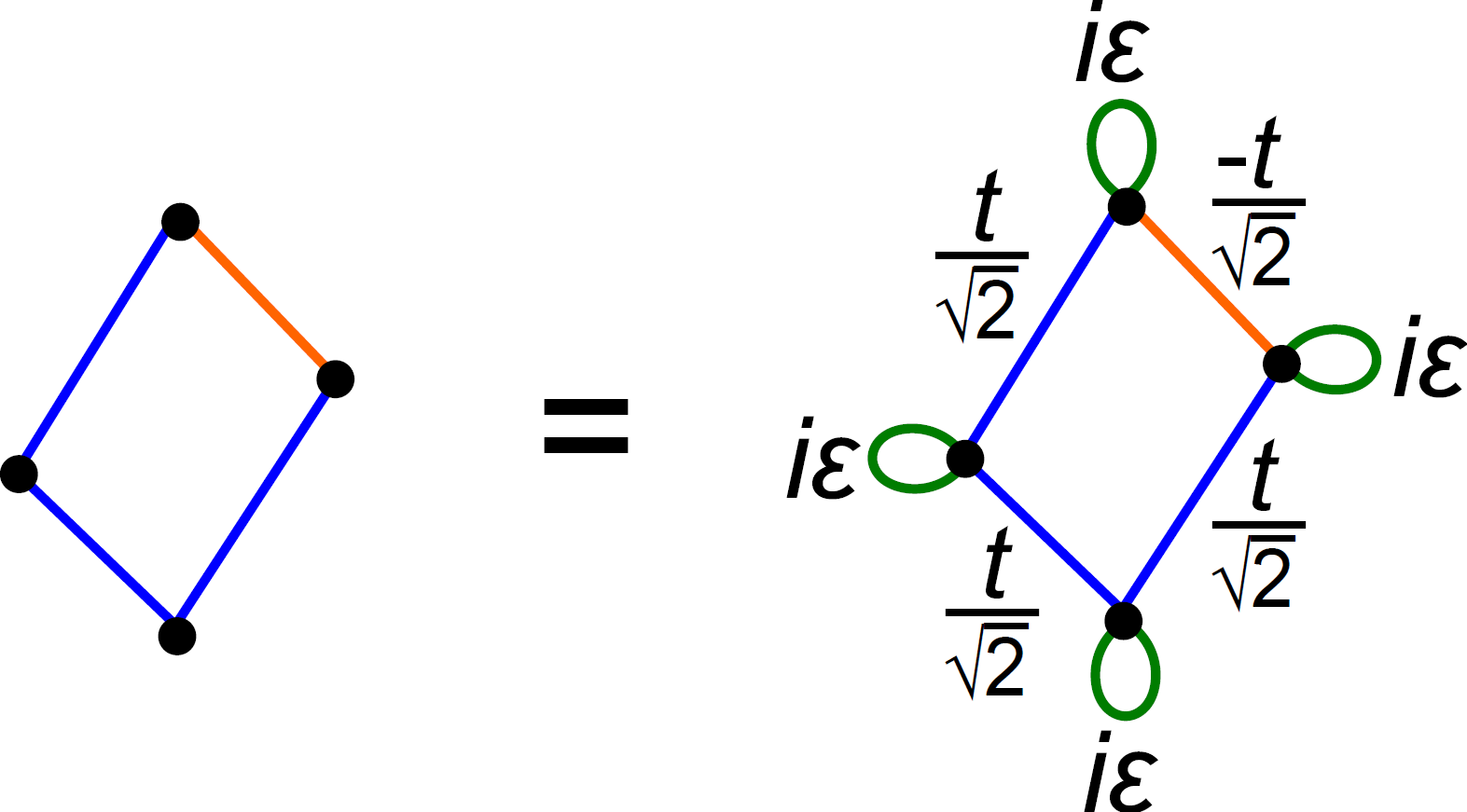}
\caption{Going from using the simplified graphical calculus description of the bilayer square lattice  with $ \mathcal{C}=2^{-1/2}$  (left)---with edge weights defined implicitly by coloring---to the full graphical calculus description (right)~\cite{Menicucci2011}. Edge weights $t$ and $\varepsilon$ are defined in text [Eqs.~\eqref{eq:tanh} and \eqref{eq:epsilon}, respectively].}
\label{fig:finitesquare}
\end{figure}

Technically, the simplified graphical calculus representations used in the majority of the figures of this  Article  are valid for both infinite- and finite-squeezing cases~\cite{Menicucci2011a}. To include finite squeezing explicitly, the full graphical calculus~\cite{Menicucci2011} must be used. To do this, simply replace the simplified disjoint square graphs in \fig{altrep}(c) by more detailed versions with self-loops and edge weights as in \fig{finitesquare}.

We note that there is a subtlety in Fig.~\ref{main:fig:setupgraph} of the main text. The state that exists at various stages \text{(a)--(d)} of the optical circuit diagram is technically not a CV cluster state, but is in fact an $\mathcal{H}$-graph state~\cite{Menicucci2011} that would have an edge weight of $- i \sinh{2r}$. However, at every stage of the diagram this state can be converted into a CV cluster state with edge weights as quoted and with the same simplified graphical representation~\cite{Menicucci2011a} by simply applying an optical phase delay of $\tfrac{\pi}{2}$ (a.k.a.\ a Fourier transform~\cite{Gu2009}) on half the qumodes (specifically, all qumodes with either even or odd frequencies). In practice, this difference is unimportant because this phase delay can be incorporated directly into the homodyne measurements acting on the final state,  and in fact, the simplified graphical calculus is  defined in Ref.~\cite{Menicucci2011a}  to represent both types of states (i.e., with or without these final Fourier transforms).

\section{Beamsplitter symmetries of a pair of two-qumode CV cluster states}
\label{sec:symmcvcs}

We use Ref.~\cite{Gabay2016} to derive the symmetries of a pair of two-qumode CV cluster states. This result is used to equate panels~(d) and~(e) of Fig.~\ref{fig:2modeproof}.

Each individual CV cluster state shown in Fig.~\ref{fig:2modeproof}(d) has a graph given by
\begin{align}
	\mat Z_1 &=
	\begin{pmatrix}
		i \sech 2r	& \tanh 2r		\\
		\tanh 2r	& i \sech 2r
	\end{pmatrix} \nonumber \\
	&= i (\sech 2r) \mat \id + (\tanh 2r) \mat \sigma_x
\end{align}
and an alternative graph representation of~\cite{Menicucci2011,Gabay2016}
\begin{align}
	\mat K_1 &\coloneqq (\mat \id+i\mat Z)(\mat \id-i\mat Z)^{-1} \nonumber \\
	&= 
	\begin{pmatrix}
		0		& i \tanh r		\\
		i \tanh r	& 0
	\end{pmatrix}
	= i(\tanh r)\mat \sigma_x\,,
\end{align}
A pair of such states (one between qumodes~$i$ and~$j$ and a separate one between qumodes~$k$ and~$l$) has the alternative graph
\begin{align}
	\mat K_2 =
	\begin{pmatrix}
		\mat K_1	& \mat 0	\\
		\mat 0	& \mat K_1
	\end{pmatrix}
	= i(\tanh r)\mat \id \otimes \mat \sigma_x\,,
\end{align}
with rows and columns ordered $(i,j,k,l)$. Note that $\otimes$ here merely indicates a matrix Kronecker product and has nothing to do with a tensor product of Hilbert spaces.

An interferometric Hamiltonian~$\tfrac 1 2 \opvec a^{\herm} \mat M \opvec a$, with ${\mat M = \mat M^\herm}$%
, generates a symmetry of the Gaussian pure state defined by~$\mat K$ if and only if $\mat M \mat K = -(\mat M \mat K)^\tp$~\cite{Gabay2016}. One choice (among many) for $\mat M$ that works for $\mat K_2$ is $\mat M = \mat \sigma_y \otimes \mat \id$. This generates a one-parameter class of symmetry operations~\cite{Gabay2016}, one example of which is
\begin{align}
	\exp\left(-i\frac \pi 4 \opvec a^{\herm} \mat M \opvec a\right)
	&= \exp \left[-\frac \pi 4 (a^\dag_i a_k + a^\dag_j a_l - \text{H.c}) \right] \nonumber \\
	&= \op B_{ik} \op B_{jl}\,.
\end{align}
Since this pair of beamsplitters is a symmetry of the pair of CV cluster states, acting with $\op B_{ik}$ alone is equivalent to acting with $ \op B_{jl}^\dag = \op B_{lj}$ alone, which is exactly the symmetry employed in Fig.~\ref{fig:2modeproof}(e).

\begin{widetext}%

\section{Derivation of the two-qumode gate}
\label{app:CZcalcs}

Here we derive the two-qumode gate [Eq.~\eqref{eq:CZgate}] implemented by using the measurement settings from Eq.~\eqref{eq:thetaCZ} in Eq.~\eqref{eq:gen2gate1}. As usual, we neglect outcome-dependent displacements and finite-squeezing effects, with discussion of these effects relegated to Sec.~\ref{sec:noise}. Before we start, we define the following abbreviations of phase shifts and squeezing on two qumodes at a time:
\begin{align}
	\op R(\theta_j,\theta_k) &\coloneqq \op R(\theta_j) \otimes \op R(\theta_k)\,, \\
	\op S(s_j,s_k) &\coloneqq \op S(s_j) \otimes \op S(s_k)\,.
\end{align}
For the chosen measurement settings [Eq.~\eqref{eq:thetaCZ}], the bottom line of Eq.~\eqref{eq:gen2gate1} (measurements of wire macronodes) gives
\begin{align}
	\op B \left[ \op V\left(\mp \frac \pi 8, \pm \frac {3\pi} {8} \right) \otimes \op V\left(\mp \frac \pi 8, \pm \frac {3\pi} {8} \right) \right] 
	= \op B%
	\choices
		{\op R\left(-\frac {3\pi} {4}, -\frac {3\pi} {4} \right)}
		{\op R\left(-\frac \pi 4, -\frac \pi 4 \right)}
	,
\end{align}
where the two cases on the right correspond to the top and bottom signs, respectively, and we omit subscripts on $\op B$ for clarity. Next, we evaluate the top line of Eq.~\eqref{eq:gen2gate1} (measurements of control macronodes), which gives
\begin{align}
	\op B &\left[\op V \left(\pm \frac \pi 4, \pm \frac \pi 4 + \phi \right) \otimes \op V \left(\pm \frac \pi 4 - \phi, \pm \frac \pi 4\right)\right]
	\nonumber \\
	&= \op B%
	\op R \left(\pm \frac \pi 4 + \frac \phi 2, \pm \frac \pi 4 - \frac \phi 2 \right)
	\op S\left(-\tan \frac \phi 2, -\tan \frac \phi 2 \right)
	\op R \left(\pm \frac \pi 4 + \frac \phi 2, \pm \frac \pi 4 - \frac \phi 2 \right)
	\nonumber \\
	&= \op B%
	\op R \left(\pm \frac \pi 4, \pm \frac \pi 4 \right)
	\op R \left(\frac \phi 2, - \frac \phi 2 \right)
	\op S\left(\tan \frac \phi 2, \tan \frac \phi 2 \right)
	\op R \left(\frac \phi 2, - \frac \phi 2 \right)
	\op R \left(\mp \frac {3\pi} {4}, \mp \frac {3\pi} {4} \right)
	\,.
\end{align}
The total gate is therefore the following product of the two lines:
\begin{align}
	\op B%
	\op R \left(\pm \frac \pi 4, \pm \frac \pi 4 \right)
	\op R \left(\frac \phi 2, - \frac \phi 2 \right)
	\op S\left(\tan \frac \phi 2, \tan \frac \phi 2 \right)
	\op R \left(\frac \phi 2, - \frac \phi 2 \right)
	\op R \left(\mp \frac {3\pi} {4}, \mp \frac {3\pi} {4} \right)
	\op B%
	\choices
		{\op R\left(-\frac {3\pi} {4}, -\frac {3\pi} {4} \right)}%
		{\op R\left(-\frac \pi 4, -\frac \pi 4 \right)}.
\end{align}
Noting that $\op B \op R ( \theta, \theta ) = \op R ( \theta, \theta ) \op B  $, 
the full gate becomes
\begin{align}
	&\quad%
	\op R \left(\pm \frac \pi 4, \pm \frac \pi 4 \right)
	\underbrace{\op B
	\op R \left(\frac \phi 2, - \frac \phi 2 \right)
	\op S\left(\tan \frac \phi 2, \tan \frac \phi 2 \right)
	\op R \left(\frac \phi 2, - \frac \phi 2 \right)
	\op B}
	_{\displaystyle \op R(\pi,0) \CZ(2\cot \phi) \op R\left(-\frac \pi 2, -\frac \pi 2 \right)} 
	\op R\left(\frac \pi 2, \frac \pi 2 \right)
	\,,
\end{align}
where we have used the Bloch-Messiah decomposition~\cite{Braunstein2005} of the $\CZ$ gate. %
This reduces the gate to its final form:
\begin{align}
	\op R \left(\mp \frac {3\pi} 4, \pm \frac \pi 4 \right)
	\CZ(2\cot \phi)
	\,.
\end{align}

\end{widetext}

\bibliographystyle{jacob_bibstyle}
\bibliography{rafref}
\end{document}

%% file: preamble.tex
\usepackage{xcolor}
\usepackage{graphicx}
\graphicspath{{Figures/}}
\usepackage{subfigure}
\usepackage[normalem]{ulem}
\usepackage{bm}

\def\({\left(}
\def\){\right)}

\def\eq#1{Eq.~(\ref{eq:#1})}

\def\fig#1{Fig.~\ref{fig:#1}}

\def\ket#1{\left|\,#1\,\right\rangle}

\makeatletter
\@ifundefined{textcolor}{}
{%
 \definecolor{BLACK}{gray}{0}
 \definecolor{WHITE}{gray}{1}
 \definecolor{RED}{rgb}{1,0,0}
 \definecolor{GREEN}{rgb}{0,.4,0}
 \definecolor{BLUE}{rgb}{0,0,1}
 \definecolor{CYAN}{cmyk}{1,0,0,0}
 \definecolor{MAGENTA}{cmyk}{0,1,0,0}
 \definecolor{YELLOW}{cmyk}{.2,.4,1,0}
 }
\makeatother

\renewcommand{\citenum}[1]{\cite{{#1}}}

\newcommand\tp{\mathrm{T}}
\newcommand\herm{\mathrm{H}}

\DeclareMathOperator{\sech}{sech}
\renewcommand{\Re}{\operatorname{Re}}
\renewcommand{\Im}{\operatorname{Im}}

\def\op#1{\hat{#1}}
\def\opvec#1{\op{\vec{#1}}}

\def\id{I}

\def\1{\mat{\id}}

\def\mat#1{\bm{\mathrm{#1}}}

\renewcommand{\vec}[1]{\bm{\mathrm{#1}}}

\def\avg#1{\left\langle {#1} \right\rangle}

\def\abs#1{\left\lvert{#1}\right\rvert}

\usepackage {fancyhdr}
\usepackage{fancyvrb}
\usepackage{braket}
\usepackage{amsthm}
\usepackage{amssymb}
\usepackage{hanging}
\usepackage{url}
\usepackage{color}
\usepackage{mathtools}

\providecommand{\abs}[1]{\left\lvert#1\right\rvert}


\usepackage{xr-hyper}
\usepackage[colorlinks=true,urlcolor=blue, linkcolor=blue,bookmarks=false]{hyperref}

\newcommand{\choices}[2]{\genfrac{\lbrace}{\rbrace}{0pt}{}{#1}{#2}}
\newcommand{\CZ}{\op C_Z}